%% file: main.tex
\documentclass[sigconf,screen,final]{acmart}
\usepackage{amsmath,amsfonts}

\AtBeginDocument{%
  }

\setcopyright{acmlicensed}
\copyrightyear{2018}
\acmYear{2018}
\acmDOI{XXXXXXX.XXXXXXX}
\acmConference[SC '25]{The International Conference for High Performance Computing, Networking, Storage and Analysis (SC '25)}{November 16--21, 2025}{St. Louis, MO}
\acmBooktitle{The International Conference for High Performance Computing, Networking,
 Storage and Analysis (SC '25), November 16--21, 2025, St. Louis, MO}
\acmISBN{978-1-4503-XXXX-X/2018/06}

\usepackage{ascmac}
\usepackage{amsthm}
\newtheorem{thm}{Theorem}
\usepackage{algorithm}
\usepackage{algpseudocode}
\usepackage{algorithmicx}

\usepackage[shortlabels]{enumitem}

\begin{document}

\title[Efficient Emulation of Matrix Multiplication using INT8 Engines]{High-Performance and Power-Efficient Emulation of\\Matrix Multiplication using INT8 Matrix Engines}


\author{Yuki Uchino}
\authornote{All authors contributed equally to this research.}
\email{yuki.uchino.fe@riken.jp}
\orcid{0000-0002-5906-6624}
\affiliation{%
  \institution{RIKEN Center for Computational Science}
  \city{Kobe}
  \state{Hyogo}
  \country{Japan}
}

\author{Katsuhisa Ozaki}
\authornotemark[1]
\email{ozaki@sic.shibaura-it.ac.jp}
\affiliation{%
 \institution{Shibaura Institute of Technology}
 \city{Saitama}
 \state{Saitama}
 \country{Japan}
}

\author{Toshiyuki Imamura}
\authornotemark[1]
\email{imamura.toshiyuki@riken.jp}
\affiliation{%
  \institution{RIKEN Center for Computational Science}
  \city{Kobe}
  \state{Hyogo}
  \country{Japan}
}

\renewcommand{\shortauthors}{Uchino, Ozaki, Imamura}

\begin{abstract}
Recent architectures integrate high-performance and power-efficient matrix engines. 
These engines demonstrate remarkable performance in low-precision matrix multiplication, which is crucial in deep learning. 
Several techniques have been proposed to emulate single- and double-precision general matrix-matrix multiplication (SGEMM and DGEMM, respectively) by leveraging such low-precision matrix engines.
In this study, we present emulation methods that significantly outperforms conventional approaches.
On a GH200 Grace Hopper Superchip, the proposed DGEMM emulation achieves a 1.4x speedup and a 43\% improvement in power efficiency compared to native DGEMM for sufficiently large problems.
The proposed SGEMM emulation achieves a 3.0x speedup and a 154\% improvement in power efficiency compared to native SGEMM for sufficiently large problems.
Furthermore, compared to conventional emulation methods, the proposed emulation achieves more than 2x higher performance and superior power efficiency.
\end{abstract}

\ccsdesc[500]{Mathematics of computing~Mathematical software performance}
\ccsdesc[500]{Computer systems organization~Multicore architectures}

\keywords{Matrix Multiplication, Emulation, High-Performance Computing, Mixed-Precision Computing, Power Efficient}

\received{20 February 2007}
\received[revised]{12 March 2009}
\received[accepted]{5 June 2009}

\maketitle

\section{Introduction}
\begin{figure}[htb]
\centering
\includegraphics[width=\hsize]{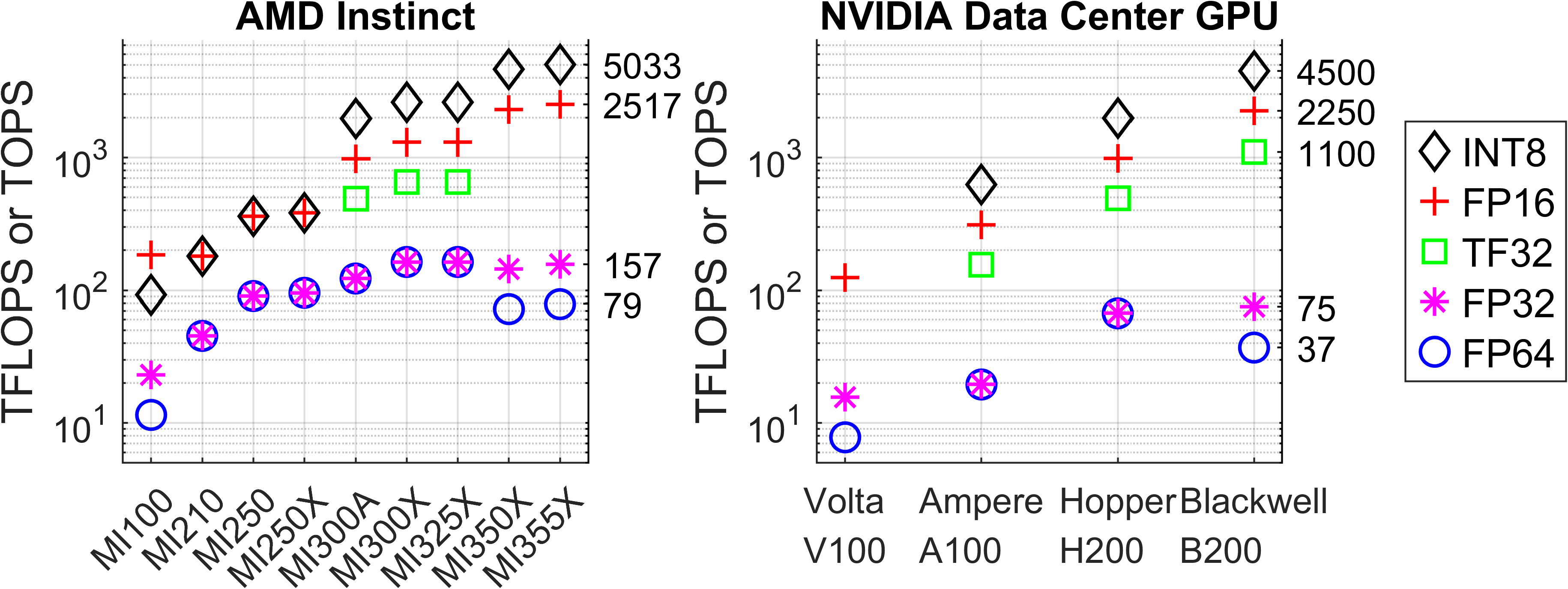}
\caption{TFLOPS and TOPS of AMD and NVIDIA GPUs for dense data}
\label{fig:specs}
\end{figure}

This paper proposes high-performance methods for emulating single- and double-precision general matrix-matrix multiplication (SGEMM and DGEMM, respectively) on modern architectures.
Recent architectures include dedicated low-precision matrix engines, such as 
NVIDIA Tensor Cores~\cite{tensorcore}, 
AMD Matrix Cores~\cite{matrixcore},
Intel AI Boost Neural Processing Units (NPU)~\cite{intelXecore},
and 
Google Cloud Tensor Processing Units (TPU)~\cite{tpu}, 
designed to accelerate the matrix operations required for deep learning. 
Figure~\ref{fig:specs} shows TFLOPS and TOPS of AMD and NVIDIA GPUs.
The performance of low-precision engines has significantly improved with each new generation. 
However, the performance increases of FP64 and FP32 are modest compared to lower precision types. 
This trend poses a challenge for the high-performance computing field, where high-precision calculations remain essential.
To overcome these challenges, various numerical mixed-precision algorithms utilizing low-precision matrix engines have been proposed. 

The present study focuses on the emulation of matrix multiplication.
Note that matrix multiplication involving tall-and-skinny or small-scale matrices is not considered in this paper, as such cases fail to fully utilize the computational capabilities of matrix engines and tend to expose performance bottlenecks in the emulation, resulting in memory-bound behavior.
Recently, we proposed Ozaki scheme II, a highly accurate matrix multiplication emulation based on the Chinese Remainder Theorem (CRT)~\cite{ozaki-scheme2}.
In the present study, we provide DGEMM and SGEMM emulation based on Ozaki scheme II using INT8 matrix engines, together with elaborately designed implementation techniques.
The code is available from~\cite{GEMMul8}.
Numerical experiment results show that our emulation is faster than native DGEMM and SGEMM for sufficiently large matrices.

The remainder of this paper is organized as follows. 
Section~\ref{sec:Related Work} introduces emulation methods proposed in the previous studies.
Section~\ref{sec:Ozaki Scheme II} gives an overview of Ozaki scheme II.
Section~\ref{sec:DGEMM and SGEMM Emulation using Matrix Engines} presents the proposed efficient implementation for emulating DGEMM and SGEMM.
Section~\ref{sec:Numerical Results} shows the numerical results of the proposed emulation.
Section~\ref{sec:Conclusion} presents the concluding remarks.

\section{Related Work}
\label{sec:Related Work}
Matrix engines can perform mixed-precision computations, accumulating the results of matrix multiplication with higher-precision than that of the input matrices.
Leveraging this property, several methods for emulating higher precision matrix multiplication have been developed.
Emulation techniques utilizing such matrix engines are emerging as an essential foundation of non-AI computations.
In the numerical experiments presented in Section~\ref{sec:Numerical Results}, we compare our emulation with these existing emulation methods.

Let $\mathbb{F}_{b}$ be the set of $b$-bit binary floating-point numbers for $b \in \mathbb{N}$.
For example, $\mathbb{F}_{32}$ and $\mathbb{F}_{64}$ are the sets of FP32 and FP64, respectively.
Let $\mathbb{Z}_{b}$ and $\mathbb{U}_b$ be the sets of $b$-bit signed and unsigned integers, respectively, for $b \in \mathbb{N}$.
cuMpSGEMM emulates SGEMM using FP16 and TF32 Tensor Cores~\cite{ootomo2022Recovering,Ootomo2023Quantum,cuMpSGEMM}.
It decomposes input matrices $A \in \mathbb{F}_{32}^{m \times k}$ and $B \in \mathbb{F}_{32}^{k \times n}$ into
$A \approx A_1 + s^{-1}A_2$ and $B \approx B_1 + s^{-1}B_2$, respectively, 
where $A_i$ and $B_i$ are represented in either FP16 or TF32 format for $i\in \{1,2\}$, and $s := 2^{-11}$ for FP16 and $s := 1$ for TF32.
After the decomposition, $AB$ is approximated as $AB \approx A_1B_1 + s^{-1}(A_1B_2 + A_2B_1)$.
cuMpSGEMM offers several computing modes.
Among them, this study uses FP16TCEC\_SCALING mode, which can emulate SGEMM using FP16 Tensor Cores without accuracy loss and with higher performance than SGEMM.
Two FP16 values cannot fully capture the mantissa of a FP32 value; therefore, error correction is applied to restore the lost mantissa precision.
As a related approach, NVIDIA cuBLAS 12.9 supports the BF16x9 algorithm for emulating SGEMM on Blackwell GPUs~\cite{bf16x9}.
It decomposes $A$ and $B$ into BF16 matrices as $A = A_1 + 2^{-8}A_2 + 2^{-16}A_3$ and $B = B_1 + 2^{-8}B_2 + 2^{-16}B_3$.
Then $AB$ is computed based on $AB = \sum_{i=1}^3 \sum_{j=1}^3 2^{-8(i+j-2)}A_iB_j$.
A similar approach using fused multiply-add (FMA) operations was proposed in~\cite{henry2019leveragingbfloat16artificialintelligence}.

The Ozaki scheme (Ozaki scheme I), a highly accurate matrix multiplication based on an error-free transformation of a matrix product~\cite{ozaki2012error,ozaki2013generalization}, can contribute to the design of emulation for higher-precision matrix multiplication using low-precision matrix engines.
For $in,out \in \mathbb{N}$ and a user-specified $d \in \mathbb{N}$, Ozaki scheme I splits two input matrices $A \in \mathbb{F}_{in}^{m \times k}$ and $B \in \mathbb{F}_{in}^{k \times n}$ into $A \approx \sum_{i=1}^d A_i$ and $B \approx \sum_{i=1}^d B_i$, respectively, where $A_i \in \mathbb{F}_{out}^{m \times k}$, $B_i \in \mathbb{F}_{out}^{k \times n}$, and no rounding error occurs in computing $A_iB_j$ for $1 \le i, j \le d$.
Then, $AB$ is approximated as $AB \approx \sum_{i+j \le d+1} A_iB_j$, involving $d(d+1)/2$ matrix multiplications.
By appropriately extracting $A_i$ and $B_j$ and applying diagonal scaling to them, $A_iB_j$ can be computed in low precision. 
In this paper, we assume that INT8 matrix engines take INT8 as an input and accumulate in INT32.
DGEMM emulation based on Ozaki scheme I using INT8 matrix engines is proposed in~\cite{ootomo2024dgemm, ozIMMU, uchino2025Performance, ozIMMU-uchino}.

\section{Ozaki Scheme II}
\label{sec:Ozaki Scheme II}
Hereafter, for any $X \in \mathbb{Z}^{m \times n}$ and $p \in \mathbb{N}$, we write 
\[
\bmod(X,p) := X - p \lfloor X/p \rfloor,\quad
\mathrm{rmod}(X,p) := X - p \cdot \mathrm{round}(X/p),
\]
where $\mathrm{round}(\cdot)$ rounds an input matrix to the nearest integers.
The methods introduced in Section~\ref{sec:Related Work} decompose $A$ and $B$ into multiple-component forms by splitting significands and transform $AB$ into a sum of the products of the corresponding components.
In contrast, Ozaki Scheme II is based on the CRT (Theorem~\ref{thm:crt}); it employs a technique that avoids splitting the input matrices.
The theorem implies that one of the solutions of a sequence of congruence equations~\eqref{eq:system} is $x = \mathrm{rmod}(\sum_{i=1}^{N} \mathcal{P}/p_i\cdot q_i\cdot y_i,\mathcal{P})$.
If $\mathcal{P}$ is sufficiently large for $|x|$, the solution becomes unique. 
Ozaki scheme II utilizes this property to emulate matrix multiplication.

\begin{thm}[Chinese Remainder Theorem]\label{thm:crt}
Let $x \in \mathbb{Z}$.
Suppose that $p_1,\dots,p_N \in \mathbb{N}_{\ge 2}$ are pairwise coprime integers and $\mathcal{P} := \prod_{1 \le i \le N}{p_i}$.
For $i=1,\dots,N$, define $q_i \in \mathbb{N}$ as modular multiplicative inverses of $\mathcal{P}/p_i$ (i.e., $\mathcal{P}/p_i \cdot q_i \equiv 1 \bmod p_i$).
Let $y_i \in \mathbb{Z}$ for $i=1,\dots,N$ be such that
\begin{equation}\label{eq:system}
\begin{cases}
    x \equiv y_1 \mod{p_1},\\
    \quad \vdots\\
    x \equiv y_N \mod{p_N}.
\end{cases}
\end{equation}
Then, it holds that
\begin{equation}\label{eq:xequiv}
  x \equiv \sum_{i=1}^{N} \frac{\mathcal{P}}{p_i} q_iy_i \mod{\mathcal{P}}.
\end{equation}
\end{thm}

Assume that $N \in \mathbb{N}$ is specified by a user.
Ozaki scheme II for computing $C \approx AB$ is constructed as follows:
\begin{description}
    \item[Step 1.] Determine pairwise coprime integers $p_1,\dots,p_N \in \mathbb{N}_{\ge 2}$ and calculate $\mathcal{P} := \prod_{1 \le i \le N}{p_i} \in \mathbb{N}$ and modular multiplicative inverses $q_1,\dots,q_N \in \mathbb{N}$ of $\mathcal{P}/p_i$.
    
    \item[Step 2.] Apply diagonal scaling and truncation to convert $A \in \mathbb{F}^{m \times k}$ and $B \in \mathbb{F}^{k \times n}$ to 
    \begin{alignat*}{2}
        A' &:= \mathrm{trunc}(\mathrm{diag}(\mu)\cdot A) \in \mathbb{Z}^{m \times k},&\quad \mu &\in \mathbb{F}^{m},\\
        B' &:= \mathrm{trunc}(B\cdot \mathrm{diag}(\nu)) \in \mathbb{Z}^{k \times n},&\quad \nu &\in \mathbb{F}^{n},
    \end{alignat*}
    respectively, where $\mathrm{trunc}(\cdot)$ rounds an input matrix toward zero and all elements of $\mu$ and $\nu$ are powers of two chosen to satisfy
    \begin{equation}\label{eq:unique}
        2\sum_{h=1}^k |a'_{ih}||b'_{hj}| < \mathcal{P}\quad \forall i,j.
    \end{equation}
    
    \item[Step 3.] Compute $C'' := A'B'$ via the CRT as
    \begin{align}
        C' &:= \sum_{i=1}^{N} \frac{\mathcal{P}}{p_i} q_i\cdot \mathrm{rmod}(A',p_i) \cdot \mathrm{rmod}(B',p_i),\label{eq:C'}\\
        C'' &:=  \mathrm{rmod}(C',\mathcal{P}).\label{eq:C''}
    \end{align}

    \item[Step 4.] Inversely scale $C'$ as
    \[
        C := \mathrm{diag}(\mu^{-1}) \cdot C'' \cdot \mathrm{diag}(\nu^{-1}).
    \]
\end{description}
$p_i$ are chosen so that no error occurs in $\mathrm{rmod}(A',p_i) \cdot \mathrm{rmod}(B',p_i)$ in~\eqref{eq:C'}.
Steps~3 performs integer matrix multiplication $A'B'$ via the CRT; $\mathrm{rmod}(A',p_i) \cdot \mathrm{rmod}(B',p_i)$ in~\eqref{eq:C'} and $C''$ in~\eqref{eq:C''} correspond to $y_i$ in~\eqref{eq:system} and $x$ in~\eqref{eq:xequiv}, respectively.
As discussed in~\cite{ozaki-scheme2}, \eqref{eq:unique} yields the uniqueness of $C''$.
In Step~2, $A$ and $B$ are transformed into $A'$ and $B'$, respectively, to ensure the uniqueness of the CRT result.
Increasing $N$ enlarges $\mathcal{P}$ in~\eqref{eq:unique}, which reduces the truncation error in Step~2. 
Therefore, the accuracy of $C \approx AB$ depends on the number of moduli.
In the following sections, we provide algorithms and coding techniques of Ozaki scheme II for emulating DGEMM and SGEMM using INT8 matrix engines.

\section{DGEMM and SGEMM Emulation}
\label{sec:DGEMM and SGEMM Emulation using Matrix Engines}
In DGEMM and SGEMM emulation, the matrix multiplication in~\eqref{eq:C'} and the other computations are performed using INT8 matrix engines and high-precision operations, respectively.
INT8 matrix engines are overwhelmingly faster than high-precision computations.
Therefore, even if the computational cost of high-precision operations is less than that of INT8 operations, their execution time remains significant and cannot be ignored.
Consequently, if Steps 2 to 4 in Section~\ref{sec:Ozaki Scheme II} are implemented in a straightforward manner, they become bottlenecks, hindering performance.
Hence, it is crucial to skillfully apply techniques such as table lookups and efficient error-free computations to avoid division and slow high-precision computations.
Taking these considerations into account, Algorithm~\ref{alg:emu} presents the proposed emulation. 
This section describes the constants and implementation details used in Algorithm~\ref{alg:emu}.

\begin{algorithm}[htb]\centering
\caption{DGEMM and SGEMM emulation via INT8\label{alg:emu}}
\begin{algorithmic}[1]
\Require $A \in \mathbb{F}_b^{m \times k}$, $B \in \mathbb{F}_b^{k \times n}$ for $b \in \{32, 64\}$, $N \in \mathbb{N}_{\ge 2}$
\State Determine scale vectors $\mu \in \mathbb{F}_b^m$ and $\nu \in \mathbb{F}_b^n$
\State $A' \gets \mathrm{trunc}(\mathrm{diag}(\mu)\cdot A) \in \mathbb{F}_b^{m \times k} \cap \mathbb{Z}^{m \times k}$\label{alg:emu:1}
\State $B' \gets \mathrm{trunc}(B\cdot \mathrm{diag}(\nu)) \in \mathbb{F}_b^{k \times n} \cap \mathbb{Z}^{k \times n}$\label{alg:emu:2}
\State $A'_i \gets \mathrm{rmod}(A',p_i) \in \mathbb{Z}_8^{m \times k}$ for $1 \le i \le N$\label{alg:emu:3}
\State $B'_i \gets \mathrm{rmod}(B',p_i) \in \mathbb{Z}_8^{k \times n}$ for $1 \le i \le N$\label{alg:emu:4}
\State $C'_i \gets A'_iB'_i \in \mathbb{Z}_{32}^{m \times n}$ for $1 \le i \le N$ using INT8 matrix engines
\State $U_i \gets \bmod(C'_i,p_i) \in \mathbb{U}_{8}^{m \times n}$ for $1 \le i \le N$\label{alg:emu:6}
\State $C'^{(1)} \gets \sum_{i=1}^N s_{i1} U_i \in \mathbb{F}_{64}^{m \times n}$
\State $C'^{(2)} \gets \sum_{i=1}^N s_{i2} U_i \in \mathbb{F}_{64}^{m \times n}$
\State $Q \gets \mathrm{round}(\mathcal{P}_{inv} \cdot C'^{(1)}) \in \mathbb{F}_{64}^{m \times n} \cap \mathbb{Z}^{m \times n}$
\State $C'' \gets ((C'^{(1)} - \mathcal{P}_1 \cdot Q) + C'^{(2)}) - \mathcal{P}_2 \cdot Q \in \mathbb{F}_{64}^{m \times n}$ using FMA
\State $C \gets \mathrm{diag}(\mu^{-1}) \cdot C'' \cdot \mathrm{diag}(\nu^{-1}) \in \mathbb{F}_{64}^{m \times n}$
\end{algorithmic}
\end{algorithm}

\subsection{Determination of Constants}
\label{subsec:Determining Constants}
When using INT8 matrix engines, the pairwise coprime integers $p_1,\dots,p_N \in \mathbb{N}_{\ge 2}$ can be fixed as
\[
p_s \in \{256, 255, 253, 251, \dots, 41, 37, 29\} \backslash \{p_t \mid t \neq s, 1 \le t \le N \}
\]
to satisfy $-128 \le (A'_s)_{ij} \le 127$ and $-128 \le (B'_s)_{ij} \le 127$.
The values $\mathrm{rmod}(A',256)$ and $\mathrm{rmod}(B',256)$ can be $128$, but casting to INT8 wraps them around to $-128$.
This is not an issue because $128 \equiv -128 \bmod 256$.
In practice, $N \le 20$ is sufficient for DGEMM emulation, while $N \le 10$ is sufficient for SGEMM emulation. 
To prevent the table size from becoming excessive, we assume $N \le 20$ in the implementation.

\begin{figure}[htb]
    \centering
    \includegraphics[width=\hsize]{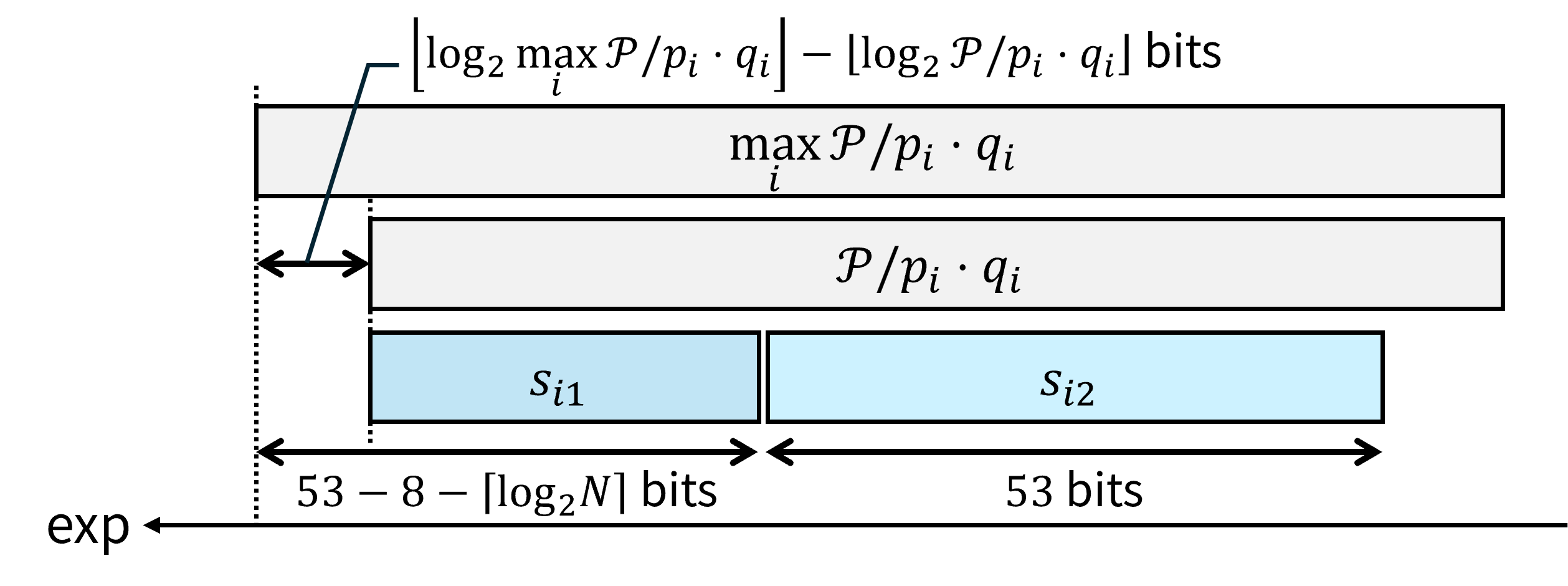}
    \caption{Image of $s_{i1}$ and $s_{i2}$}
    \label{fig:img_si1si2}
\end{figure}

We define $\mathrm{double}(\cdot)$ and $\mathrm{single}(\cdot)$ to denote rounding to nearest double- and single-precision floating-point numbers, respectively.
Since $p_i$ is fixed, both $\mathcal{P}$ and $q_i$ are known in advance.
Therefore, the implementation constructs a table of $p_i$, $\mathcal{P}$, and $\mathcal{P}/p_i \cdot q_i$ for each $N$. 
In DGEMM emulation, accumulation in Step~3 requires arithmetic with precision higher than double precision; thus, $\mathcal{P}$ is stored as a double-double number:
\[
\mathcal{P}_1 := \mathrm{double}(\mathcal{P}) \in \mathbb{F}_{64},\quad
\mathcal{P}_2 := \mathrm{double}(\mathcal{P} - \mathcal{P}_1) \in \mathbb{F}_{64},
\]
and $\mathcal{P}/p_i \cdot q_i$ is approximated using two double-precision floating-point numbers $s_{i1},s_{i2} \in \mathbb{F}_{64}$ in the form
\begin{equation}\label{eq:s1s2}
\frac{\mathcal{P}}{p_i} \cdot q_i \approx s_{i1} + s_{i2},
\end{equation}
where this is not a double-double number.
As discussed later (see Section~\ref{subsec:Accumulation}), to reduce the computational cost of the accumulation, $s_{i1}$ retains only the upper $\beta_i$ bits of $\mathcal{P}/p_i \cdot q_i$ for
\[
\beta_i := 53 - 8 - \lceil \log_2 N \rceil + \left\lfloor \log_2 \frac{\mathcal{P}}{p_i}q_i \right\rfloor - \left\lfloor \log_2 \max_{1 \le j \le N} \frac{\mathcal{P}}{p_j}q_j \right\rfloor,
\]
while $s_{i2}$ holds the upper $53$ bits of the remainder.
Figure~\ref{fig:img_si1si2} shows an image of $s_{i1}$ and $s_{i2}$.

In SGEMM emulation, we prepare $\mathcal{P}$ and $\mathcal{P}/p_i \cdot q_i$ as double-precision floating-point numbers:
\[
\mathcal{P}_1 := \mathrm{double}(\mathcal{P}) \in \mathbb{F}_{64},\quad
s_{i1} := \mathrm{double}\left(\frac{\mathcal{P}}{p_i} \cdot q_i\right) \in \mathbb{F}_{64},
\]
and we set $\mathcal{P}_2 = s_{i2} = 0$.
In addition, for later use, 
\begin{align*}
    \mathcal{P}_{inv} &:= \mathrm{double}(\mathcal{P}^{-1}) \in \mathbb{F}_{64},\\
    \mathcal{P}'_{fast} &:= \mathrm{single}(\log_2 (\mathcal{P}-1)-1.5) \in \mathbb{F}_{32},\\
    \mathcal{P}'_{accu} &:= \mathrm{single}(\log_2 (\mathcal{P}-1)-0.5) \in \mathbb{F}_{32},\\
    pinv_i^{(64)} &:= \mathrm{double}(1/p_i) \in \mathbb{F}_{64},\\
    pinv_i^{(32)} &:= \mathrm{single}(1/p_i) \in \mathbb{F}_{32},\\
    pinv_i' &:= \lfloor 2^{32}/p_i - 1 \rfloor \in \mathbb{Z}_{32}
\end{align*}
are also prepared.

\subsection{Conversion to INT8}
\label{subsec:Conversion to INT8}
We provide two computing modes: fast mode and accurate mode.
In fast mode, the scale vectors $\mu \in \mathbb{F}^{m}_b$ and $\nu \in \mathbb{F}^n_b$ are determined using the Cauchy--Schwarz inequality as
\begin{equation}\begin{aligned}
2\sum_{h=1}^k |a'_{ih}||b'_{hj}|
&\le 2\mu_i^{-1}\left\|a_{i,:}\right\|_2\left\|b_{:,j}\right\|_2\nu_j^{-1}
< \mathcal{P}.
\end{aligned}\label{eq:the Cauchy--Schwarz inequality}\end{equation}
To satisfy this, $\mu$ and $\nu$ are set as
\begin{align*}
\mu_i &:= 2^{\left\lfloor \mathcal{P}'_{fast}-\max\left(1,\ 0.51 \cdot \log_2 \sum_{j=1}^k a_{ij}^2 \right) \right\rfloor - \left\lfloor \log_2 \max_{1 \le h \le k}|a_{ih}| \right\rfloor},\\
\nu_j &:= 2^{\left\lfloor \mathcal{P}'_{fast}-\max\left(1,\ 0.51 \cdot \log_2 \sum_{i=1}^k b_{ij}^2 \right) \right\rfloor - \left\lfloor \log_2 \max_{1 \le h \le k}|b_{hj}| \right\rfloor},
\end{align*}
where $\sum_{j=1}^k a_{ij}^2$ and $\sum_{i=1}^k b_{ij}^2$ should be computed using floating-point arithmetic in round-up mode.
In accurate mode, the scale vectors $\mu' \in \mathbb{F}^{m}_b$ and $\nu' \in \mathbb{F}^n_b$ are determined as 
\[
\mu'_i := 2^{5 - \left\lfloor \log_2 \max_{1 \le h \le k}|a_{ih}| \right\rfloor},\quad
\nu'_j := 2^{5 - \left\lfloor \log_2 \max_{1 \le h \le k}|b_{hj}| \right\rfloor},
\]
and then $A$ and $B$ are converted to INT8 matrices $\overline{A} \in \mathbb{Z}^{m \times k}$ and $\overline{B} \in \mathbb{Z}^{k \times n}$, respectively, as follows:
\[
\overline{a}_{ij} := \lceil\mu'_i\cdot |a_{ij}|\rceil \le 2^{7}-1,\quad
\overline{b}_{ij} := \lceil |b_{ij}|\cdot \nu'_j \rceil \le 2^{7}-1.
\]
Then, the scale vectors $\mu$ and $\nu$ can be determined to satisfy
\begin{align*}
2\sum_{h=1}^k |a'_{ih}||b'_{hj}|
&\le 2\mu_i^{-1}\left(\sum_{h=1}^k |a_{ih}||b_{hj}|\right)\nu_j^{-1}\\
&\le 2\mu_i^{-1}\mu'^{-1}_i \left(\overline{A} \cdot \overline{B} \right)_{ij} \nu'^{-1}_j\nu_j^{-1}
< \mathcal{P}.
\end{align*}
This mode computes $\overline{C} := \overline{A}\cdot \overline{B}$ using INT8 matrix engines.
Then $\mu$ and $\nu$ are set as
\begin{align*}
\mu_i &:= \mu'_i \cdot 2^{\left\lfloor \mathcal{P}'_{accu} - 0.51 \cdot \log_2(\max_{1 \le h \le n}\overline{c}_{ih}) \right\rfloor},\\
\nu_j &:= \nu'_j \cdot 2^{\left\lfloor \mathcal{P}'_{accu} - 0.51 \cdot \log_2(\max_{1 \le h \le m}\overline{c}_{hj}) \right\rfloor}.
\end{align*}

In terms of computation speed, fast mode is quicker than accurate mode because accurate mode requires matrix multiplication $\overline{A}\cdot \overline{B}$.
However, when the accuracy of results is required, accurate mode should be used.
It outputs more accurate results than those of fast mode because the Cauchy--Schwarz inequality overestimates the upper bound of $\sum_{h=1}^k |a_{ih}||b_{hj}|$ in~\eqref{eq:the Cauchy--Schwarz inequality}, whereas accurate mode estimates the upper bound by direct matrix multiplication.

After the scale values have been determined, $A$ and $B$ are converted to INT8 matrices 
$A'_1,\dots,A'_N \in \mathbb{Z}^{m \times k}$ and 
$B'_1,\dots,B'_N \in \mathbb{Z}^{k \times n}$, 
respectively, as in the lines~\ref{alg:emu:3} and \ref{alg:emu:4} of Algorithm~\ref{alg:emu}.
Since the built-in modular arithmetic function $\mathrm{fmod}(\cdot)$ is not efficient, we implement $\mathrm{rmod}(x, p_i)$ for $x \in \mathbb{F}_b$ for $b \in \{32,64\}$ as follows to achieve better performance:
\begin{enumerate}
    \item $y \gets \mathrm{single}(\mathrm{fma}(\mathrm{round}(x \cdot pinv_i^{(b)}),-p_i,x))$,
    \item if $N \ge N_1$, $y \gets \mathrm{fma}(\mathrm{round}(y \cdot pinv_i^{(32)}),-p_i,y)$,
    \item if $N \ge N_2$, $y \gets \mathrm{fma}(\mathrm{round}(y \cdot pinv_i^{(32)}),-p_i,y)$,
\end{enumerate}
where the function $\mathrm{fma}(x,y,z)$ calculates $xy-z$ using FMA and $(N_1,N_2) = (13,19)$ and $(N_1,N_2) = (5,11)$ for $b = 64$ and $b = 32$, respectively.
This $\mathrm{rmod}(x, p_i)$ works well for $N \le 18$ when $b = 32$ and for $N \le 20$ when $b = 64$.

\subsection{Accumulation}
\label{subsec:Accumulation}
For $i=1,\dots,N$, INT32 matrices $C'_i := A'_i \cdot B'_i$ are calculated using INT8 matrix engine.
Performing FP64 accumulation in each matrix multiplication step to compute $C' := \sum_{i=1}^{N} \mathcal{P}/p_i \cdot q_i\cdot C'_i$ creates a performance bottleneck. 
Thus, we convert $C'_i$ to UINT8 matrices $U_i$ as in the line~\ref{alg:emu:6} of Algorithm~\ref{alg:emu}, then accumulate all $U'_i$ in a single kernel invocation in the next stage.
We use $\bmod(C'_i, p_i)$ instead of $\mathrm{rmod}(C'_i, p_i)$ because integer arithmetic performs truncation rather than round-to-nearest.
Because integer modulo operation \% is relatively slow, we implemented $\bmod(x, p_i)$ for $x \in \mathbb{Z}_{32}$ as follows:
\begin{enumerate}
    \item $y \gets x - \mathrm{\_\_mulhi}(x, pinv_i')\cdot p_i$,
    \item $y \gets y - (y \ge p_i)\cdot p_i$,
    \item $y \gets y + (y < 0)\cdot p_i$,
\end{enumerate}
where $\mathrm{\_\_mulhi}$ calculates the most significant 32 bits of the product.

We assume that $k <= 2^{17}$.
For $k < 2^{17}$, $A'_i \cdot B'_i$ can be calculated without error because all elements of the product of $A'_i \cdot B'_i$ are less than $2^{31}$.
For $k = 2^{17}$, $p_1 = 256$ implies $(A'_1 \cdot B'_1)_{ij} \le 2^{31}$.
Even when $(A'_1 \cdot B'_1)_{ij} = 2^{31}$, wraparound occurs and we obtain $(C'_1)_{ij}=-2^{31}$.
Since $\mathrm{rmod}(-2^{31}, 256) = \mathrm{rmod}(2^{31}, 256) = 0$, no error is introduced for computing $(U_1)_{ij}$.
If we have case $k > 2^{17}$, it is possible to apply block matrix multiplication.

Using $U_i$, $C'$ in \eqref{eq:C'} can be represented as $C' := \sum_{i=1}^{N} \mathcal{P}/p_i \cdot q_i\cdot U_i$.
We compute $C' \approx C'^{(1)} + C'^{(2)}$ as
\[
C'^{(1)} \gets \sum_{i=1}^{N}s_{i1}U_i,\quad
C'^{(2)} \gets \sum_{i=1}^{N}s_{i2}U_i,
\]
where $s_{i1}$ and $s_{i2}$ are defined as in \eqref{eq:s1s2} and no rounding error occurs in $\sum_{i=1}^{N}s_{i1}U_i$ in FP64.
While direct accumulation of $C'_i \in \mathbb{Z}^{m\times n}_{32}$ lacks sufficient accuracy, using $U_i \in \mathbb{U}^{m \times n}_{8}$ enables sufficiently accurate summation.
Then, we compute 
\begin{equation}
    Q \gets \mathrm{round}(\mathcal{P}_{inv} \cdot C'^{(1)}) = \mathrm{round}(C'/\mathcal{P})\label{eq:Q}
\end{equation}
and $C''$ in \eqref{eq:C''} as
\[
    C'' \gets \mathrm{fma}(-\mathcal{P}_2,Q,\mathrm{fma}(-\mathcal{P}_1,Q,C'^{(1)}) + C'^{(2)}) \approx C' - \mathcal{P}Q.
\]
A rigorous error analysis of the proposed method, which will also establish the validity of~\eqref{eq:Q}, will be pursued as future work.

\section{Numerical Results}
\label{sec:Numerical Results}
Numerical experiments in this paper were conducted on 
an NVIDIA GeForce RTX 5080 Blackwell GPU with an AMD Ryzen 9 7950X CPU, 
an NVIDIA Ampere A100 SXM4 GPU with an AMD EPYC 7713 CPU, and 
an NVIDIA GH200 Grace Hopper Superchip.
All code was compiled with NVIDIA CUDA Toolkit 12.9.86 and gcc 11.5.0.
Here, we compare the following methods:
\begin{itemize}[leftmargin=*,widest=ozIMMU\_EF-Ss]
    \item[DGEMM:] Native DGEMM using cublasGemmEx.
    \item[SGEMM:] Native SGEMM using cublasGemmEx.
    \item[TF32GEMM:] TF32 matrix multiplication using cublasGemmEx\\
    with CUBLAS\_COMPUTE\_32F\_FAST\_TF32.
    \item[BF16x9:] SGEMM using cublasGemmEx with\\
    CUBLAS\_COMPUTE\_32F\_EMULATED\_16BFX9.
    \item[cuMpSGEMM:] SGEMM using cuMpSGEMM in\\FP16TCEC\_SCALING mode~\cite{cuMpSGEMM}.
    \item[ozIMMU\_EF-$S$:] DGEMM using Ozaki scheme I with $S$ slices~\cite{ozIMMU-uchino}.
    \item[OS II-fast-$N$:] Ozaki scheme II in fast mode with $N$ moduli.
    \item[OS II-accu-$N$:] Ozaki scheme II in accurate mode with $N$ moduli.
\end{itemize}
ozIMMU\_EF-$S$ was not developed for emulating SGEMM; it was not compared in the numerical experiments of SGEMM emulation.
$A \in \mathbb{F}_{b}^{m \times k}$ and $B \in \mathbb{F}^{k \times n}_b$ for $b \in \{32,64\}$ were generated as
\[
    a_{ij},b_{ij} \gets (\mathrm{rand}-0.5)\cdot \exp(\phi\cdot \mathrm{randn}),
\]
where 
$\phi \in \mathbb{F}_b$ controls the exponent distribution, 
$\mathrm{rand} \in (0,1] \subset \mathbb{F}_b$ is a uniformly distributed random number, and 
$\mathrm{randn} \in \mathbb{F}_b$ is drawn from the standard normal distribution.
Both $\mathrm{rand}$ and $\mathrm{randn}$ are generated using the cuRAND API with a fixed seed value. 
Empirically, the exponent distribution for matrix multiplication in HPL~\cite{dongarra2003linpack} is comparable to $\phi = 0.5$.

\begin{figure*}[htb]
  \centering
  \includegraphics[width=\linewidth]{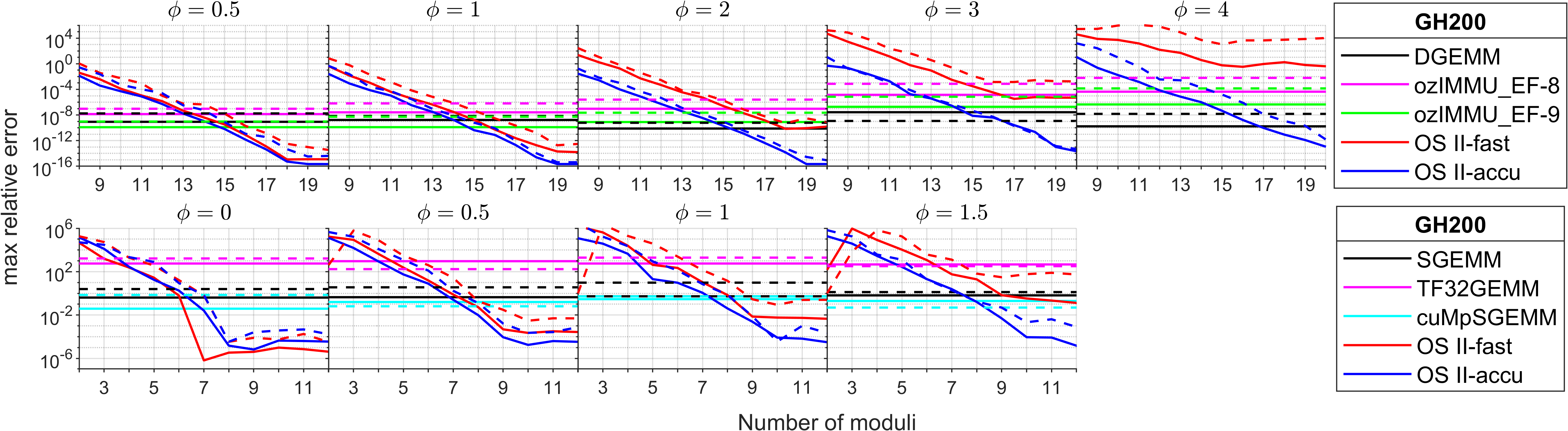}
  \caption{Accuracy of DGEMM (top) and SGEMM (bottom) emulation for $m=n=1024$ on GH200. Solid lines represent results for $k=1024$, and dashed lines for $k=16384$.\label{fig:accuracy}}
\end{figure*}

\subsection{Accuracy}
\label{subsec:Accuracy}
Figure~\ref{fig:accuracy} shows the accuracy of DGEMM emulation and SGEMM emulation on GH200.
Similar results were obtained on A100 and RTX 5080.
For $\phi = 0.5$, OS II-fast-$14$ yielded results with slightly lower accuracy than that of DGEMM, whereas OS II-accu-$14$ achieved accuracy comparable to that of DGEMM.
OS II-fast-$15$ produced results with accuracy on par with DGEMM, while OS II-accu-$15$ provided slightly higher accuracy than DGEMM.
These results imply that HPL can employ emulation with $14$ or $15$ moduli.
For larger $\phi$, the limiting accuracy of OS II-fast-$N$ got worse as $\phi$ increased.
The overestimation of the upper bound of $\sum_{h=1}^k |a'_{ih}||b'_{hj}|$ in \eqref{eq:the Cauchy--Schwarz inequality} yields significantly small shift values $\mu$ and $\nu$; the truncation errors in the lines~\ref{alg:emu:1} and \ref{alg:emu:2} of Algorithm~\ref{alg:emu} become huge in fast mode.
In contrast, accurate mode estimates the upper bound by direct matrix multiplication.
Therefore, OS II-accu-$N$ exhibits smaller truncation errors compared to those of OS II-fast-$N$, allowing it to achieve sufficient accuracy with $N \leq 17$ even for $\phi = 4$.

On RTX 5080, SGEMM and BF16x9 exhibited equivalent accuracy.
For $\phi \in \{0.5,1,1.5\}$, OS II-fast-$2$ yields $A' = O$ and $B' = O$ due to overestimation in~\eqref{eq:the Cauchy--Schwarz inequality}.
For $\phi \le 1$, OS II-fast-$N$ with $N \in \{7, 8\}$ returned results with SGEMM-level accuracy.
For $\phi = 1.5$, the limiting accuracy of OS II-fast-$N$ was around the accuracy of SGEMM; OS II-fast-$9$ achieved SGEMM-level accuracy for $k = 1024$.
For $\phi \le 1.5$, OS II-accu-$N$ with $N \in \{6, 7, 8\}$ achieved accuracy comparable to that of SGEMM.
In addition, OS II-fast-$N$ with $N \in \{4, 5, 6, 7\}$ and OS II-accu-$N$ with $N \in \{4, 5\}$ achieved TF32-level accuracy.

\subsection{Throughput Performance}
\label{subsec:Throughput Performance}
\begin{figure*}[htb]
  \centering
  \includegraphics[width=\hsize]{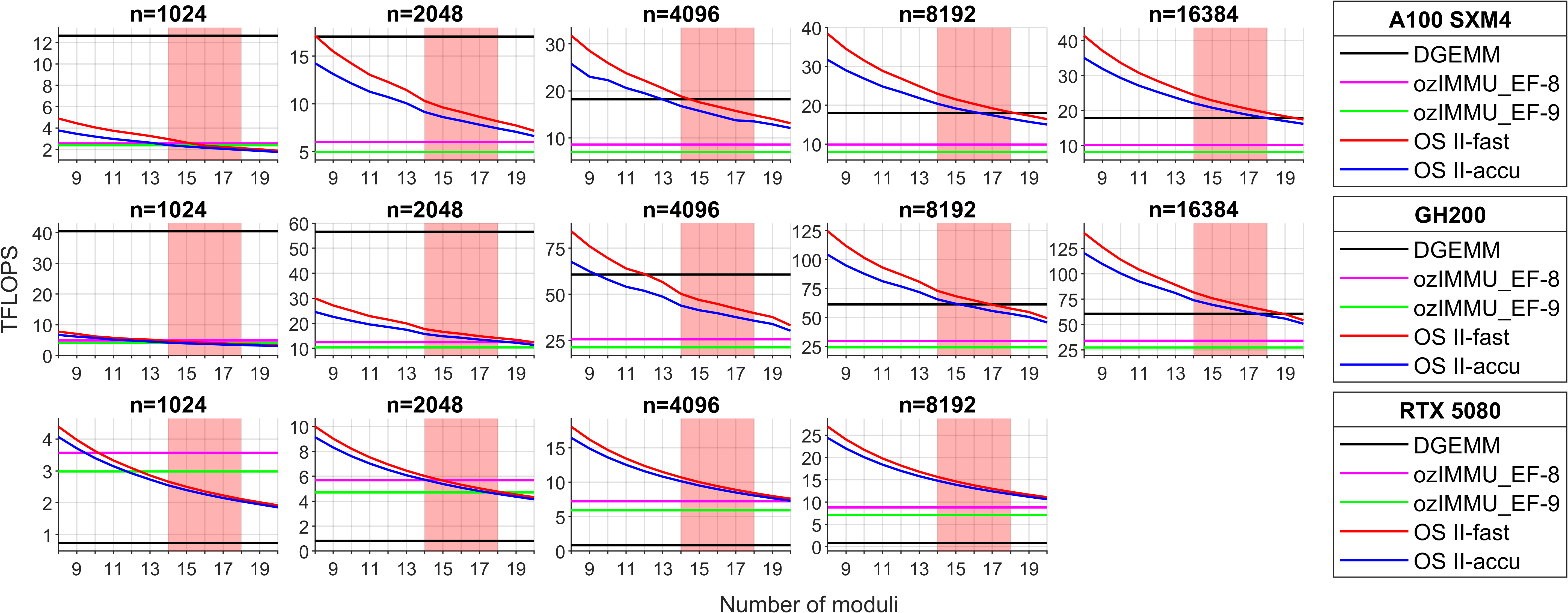}
  \caption{Throughput performance of DGEMM emulation on A100 (top), GH200 (middle), and RTX 5080 (bottom)\label{tab:performance-DGEMM}}
\end{figure*}

Figure~\ref{tab:performance-DGEMM} shows the throughput performance of DGEMM and its emulation for $m = n = k$.
Note that the scales of the vertical axes are not uniform.
According to Figure~\ref{fig:accuracy}, $N \in \{14,15,16, 17\}$ yields a DGEMM-level accuracy.
On RTX 5080, emulation methods were much faster than native DGEMM even for $n = 1024$ because FP64 is much slower than INT8 Tensor Cores, as shown in Figure~\ref{fig:specs}.
OS II-fast-$14$ and OS II-accu-$14$ achieved 18.5x and 17.5x speedups, respectively, compared to DGEMM for $n = 8192$.
For DGEMM-level results for $n \le 2048$, ozIMMU\_EF-$S$ was faster than OS II-fast-$N$ and OS II-accu-$N$ except for OS II-fast-$14$ for $n = 2048$ because computation in FP64 is the majority of the computation of Ozaki scheme II for small problems.
For sufficiently large $n$, OS II-fast-$N$ and OS II-accu-$N$ were faster than ozIMMU\_EF-$S$.
On A100 and GH200, FP64 is not as slow as RTX 5080; therefore, OS II-fast-$N$ and OS II-accu-$N$ were faster than or comparable to ozIMMU\_EF-$S$ even for $n \le 2048$.
For $n \ge 8192$, OS II-fast-$N$ and OS II-accu-$N$ outperformed DGEMM because the overhead of computation in FP64 becomes relatively small.
For $n=16384$, OS II-fast-14 achieved 24.5 TFLOPS on A100 and 81.6 TFLOPS on GH200, making it approximately 1.4x faster than DGEMM.

\begin{figure*}[htb]
  \centering
  \includegraphics[width=\hsize]{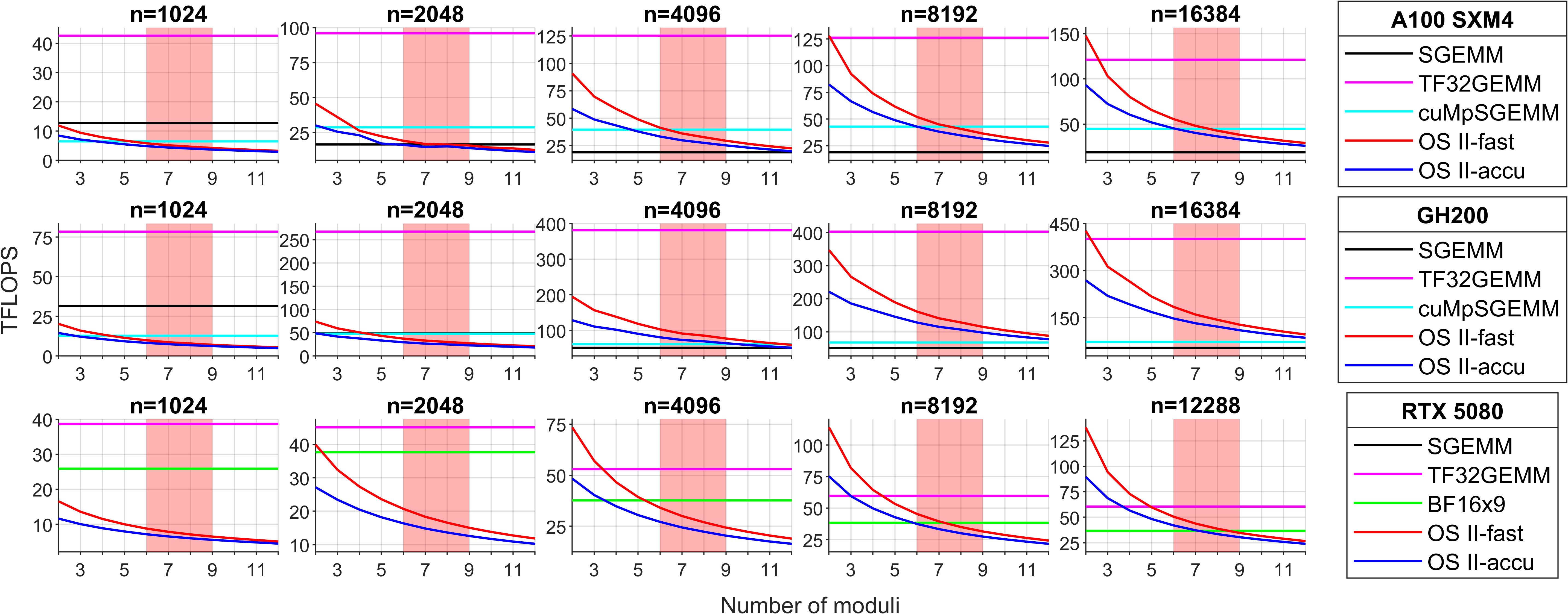}
  \caption{Throughput performance of SGEMM emulation on A100 (top), GH200 (middle), and RTX 5080 (bottom)\label{tab:performance-SGEMM}}
\end{figure*}

Figure~\ref{tab:performance-SGEMM} shows the throughput performance of SGEMM and its emulation for $m = n = k$.
The implementation of cuMpSGEMM was made from scratch using the NVIDIA WMMA API (Tensor Core device API) and optimized for A100, which may result in suboptimal performance on RTX 5080 and GH200.
While it was not functional on RTX 5080, it outperformed SGEMM on GH200.
The results of SGEMM and BF16x9 were comparable.
For SGEMM-level results on RTX 5080, OS II-fast-$N$ with $N \in \{6,7,8\}$ was faster than SGEMM and BF16x9 for $n=12288$.
On A100, as described above, cuMpSGEMM was optimized; however, OS II-fast-$6$ and $7$ were faster than or comparable to cuMpSGEMM for $n \ge 4096$.
For $n=16384$ on GH200, OS II-accu-$N$ with $N \in \{6,7,8\}$ and OS II-fast-$N$ with $N \in \{7,8,9\}$ achieved 122--148 TFLOPS and 128--160 TFLOPS, respectively, indicating that Ozaki scheme II achieved a 2.3--3.0x speedup compared to SGEMM.
In addition, Ozaki scheme II demonstrated performance between those of SGEMM and TF32GEMM. 
As shown in Figure~\ref{fig:accuracy}, Ozaki scheme II achieved accuracy between those of SGEMM and TF32GEMM, indicating that it can serve as an intermediate-precision approach between FP32 and TF32.

\subsection{Time Breakdown}
\label{subsec:Time Breakdown}
\begin{figure*}[htb]
  \centering
  \begin{minipage}{.49\hsize}
  \includegraphics[width=\hsize]{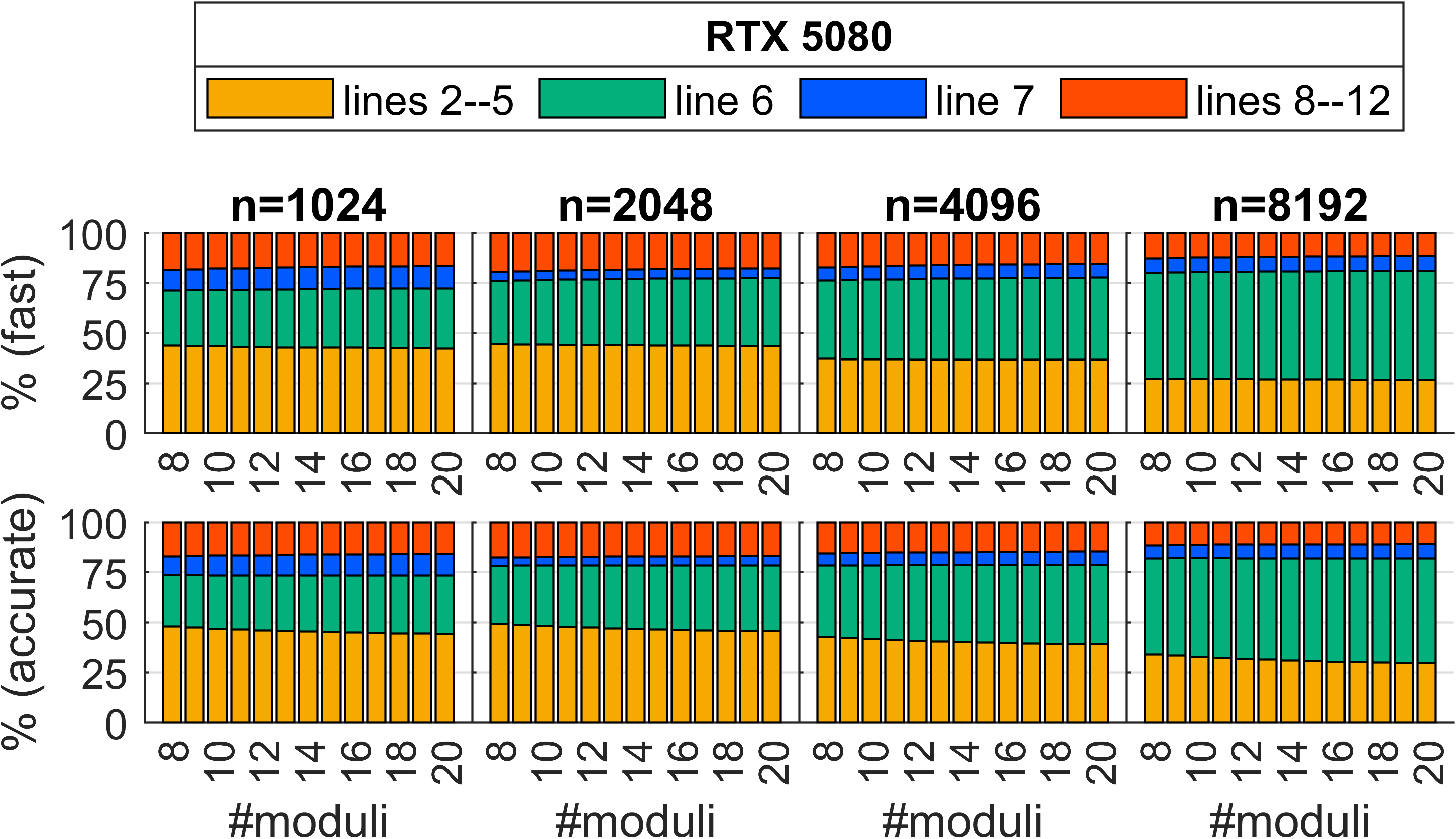}
  \end{minipage}
  \begin{minipage}{.49\hsize}
  \includegraphics[width=\hsize]{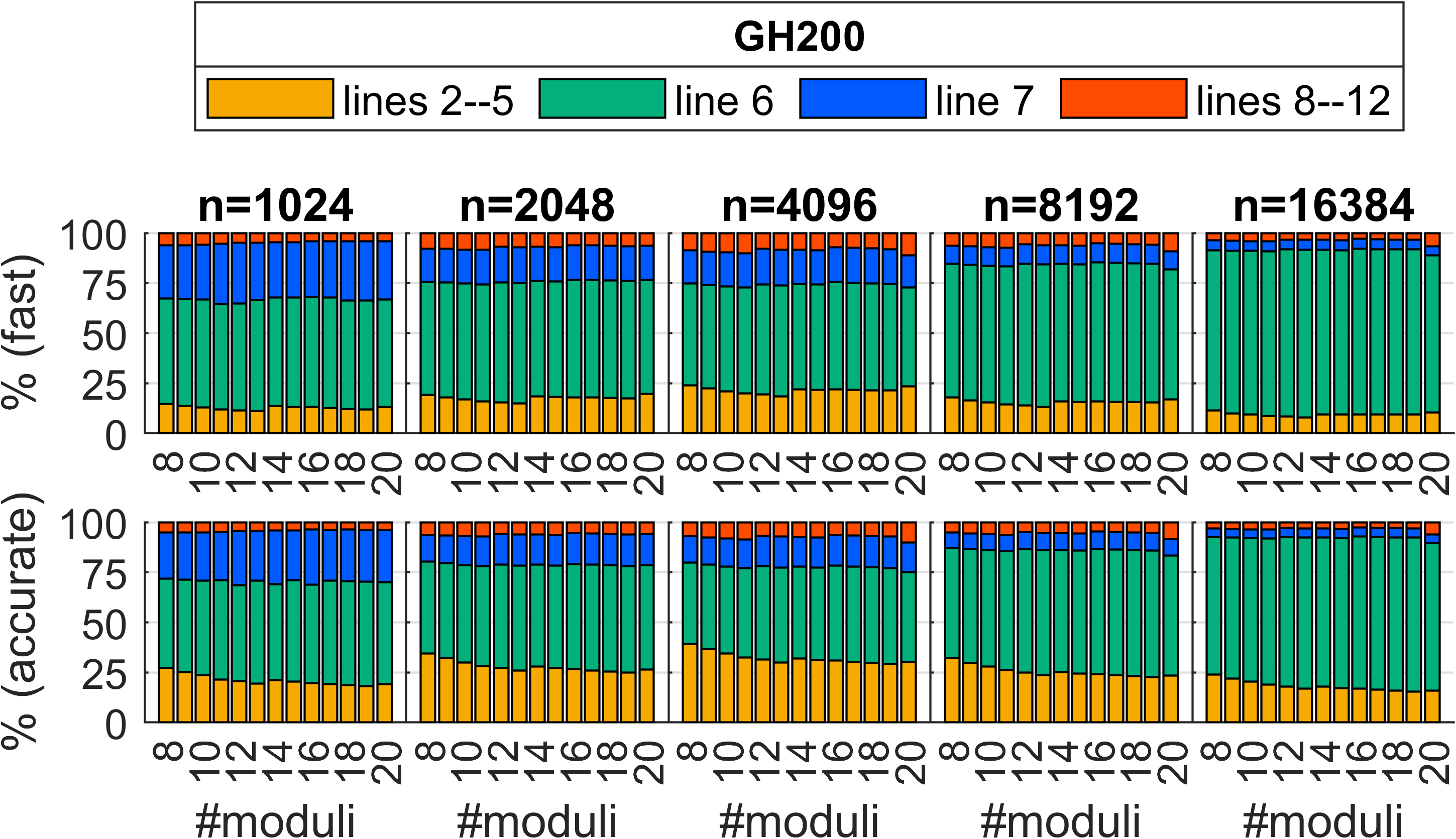}
  \end{minipage}
  \caption{Time breakdown of DGEMM emulation in fast (top) and accurate (bottom) modes on RTX 5080 (left) and GH200 (right)\label{fig:time-breakdown-DGEMM}}
\end{figure*}

\begin{figure*}[htb]
  \centering
  \begin{minipage}{.49\hsize}
  \includegraphics[width=\hsize]{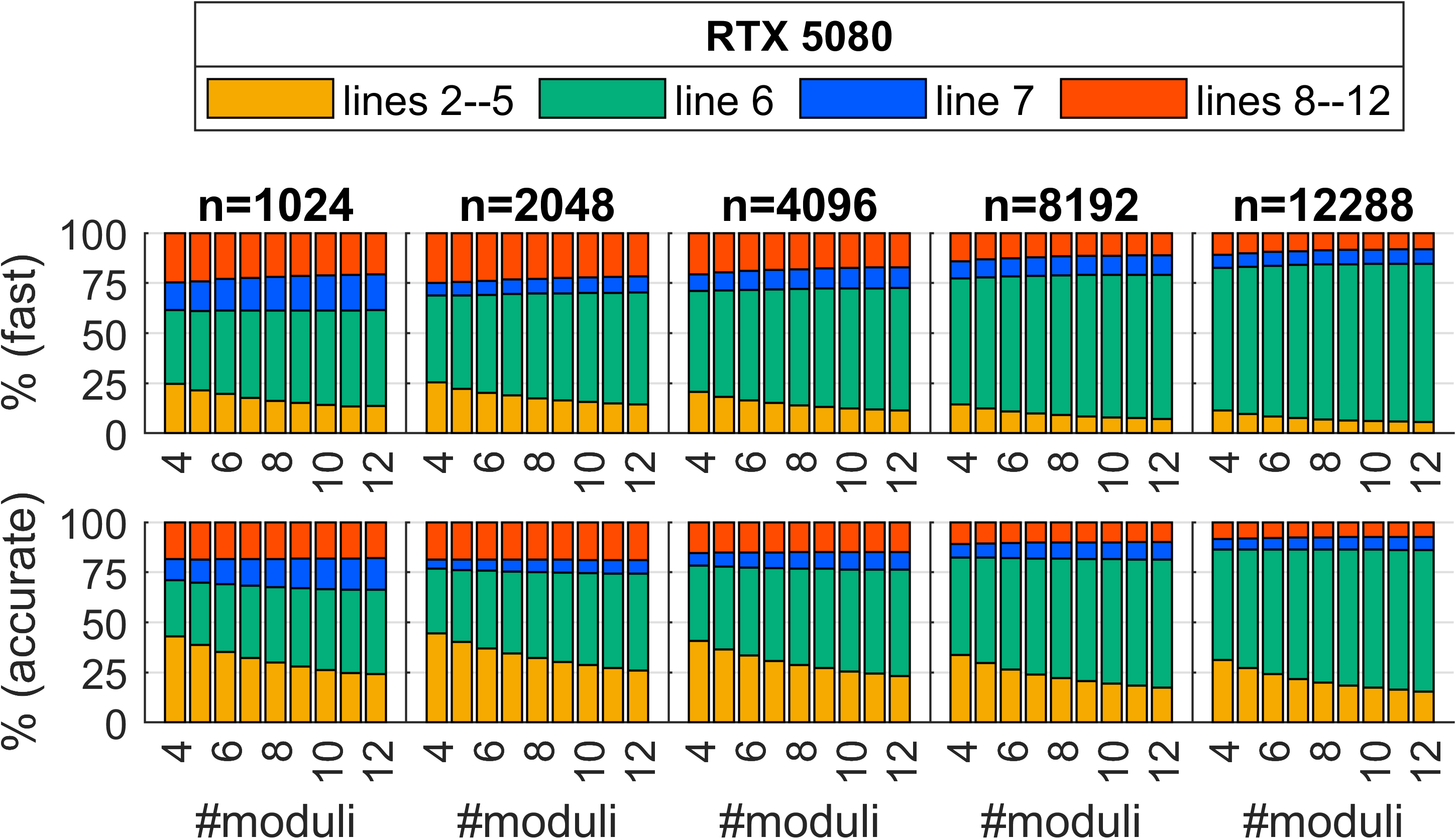}
  \end{minipage}
  \begin{minipage}{.49\hsize}
  \includegraphics[width=\hsize]{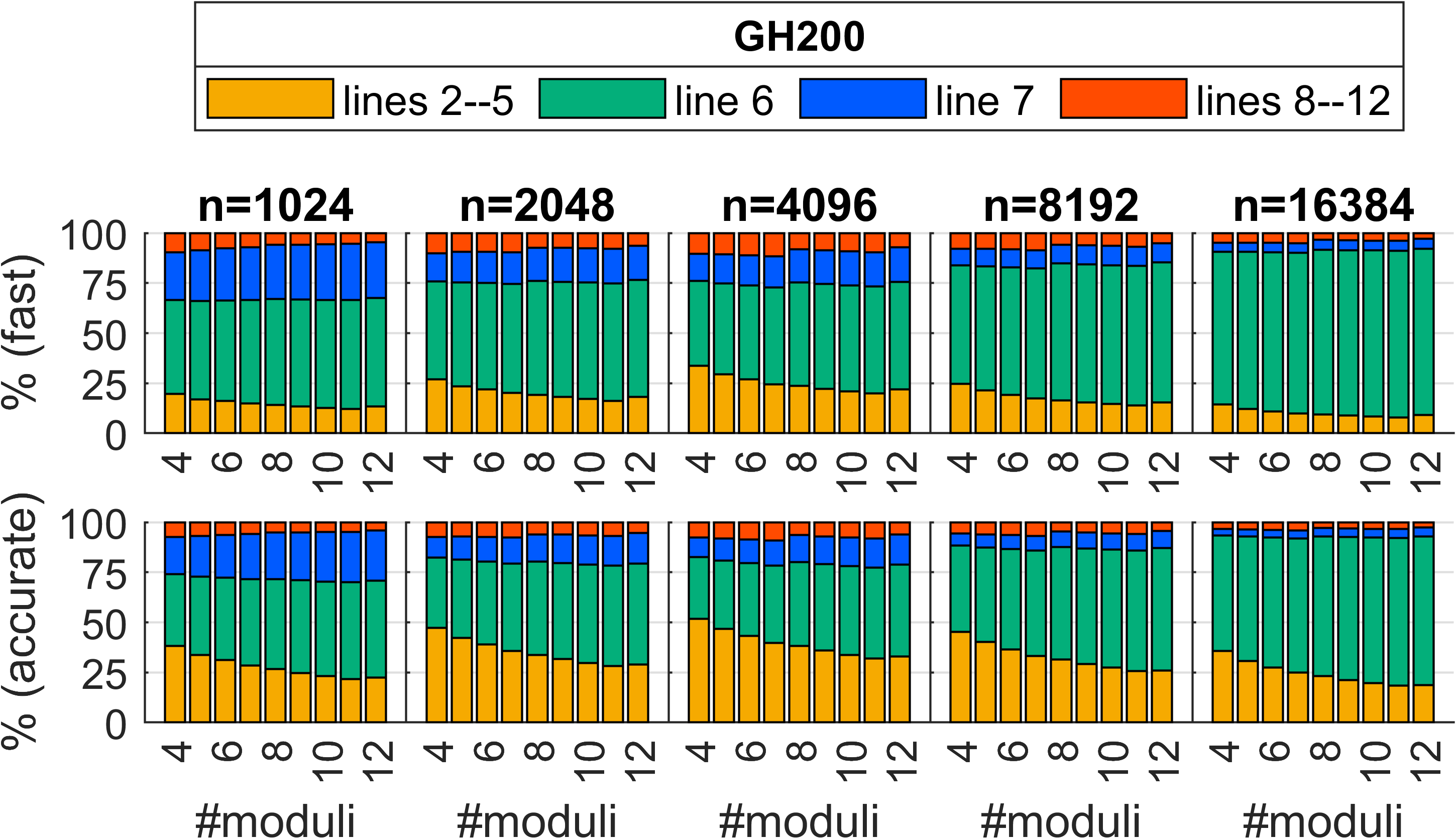}
  \end{minipage}
  \caption{Time breakdown of SGEMM emulation in fast (top) and accurate (bottom) modes on RTX 5080 (left) and GH200 (right)\label{fig:time-breakdown-SGEMM}}
\end{figure*}

Figures~\ref{fig:time-breakdown-DGEMM} and \ref{fig:time-breakdown-SGEMM} show the time breakdown of emulation of DGEMM and SGEMM on RTX 5080 and GH200 for $m = n = k$.
The results for A100 were similar to those for GH200.
Legend labels in the figures correspond to the respective lines in Algorithm~\ref{alg:emu}.
The conversion of input matrices in accurate mode includes matrix multiplication and accounts more computation time than that of fast mode.
For DGEMM emulation on RTX 5080, due to low-performance FP64, even for $n=8192$, non-matrix multiplication components accounted for around 50\% of the entire computation time.
For SGEMM emulation on RTX 5080, FP32 is 64x faster than FP64; thus, the conversion of input matrices is accelerated compared to that of DGEMM emulation.
On A100 and GH200, for sufficiently large $n$, matrix multiplication is the major computation.
As $n$ increases, computations except for matrix multiplication gradually become negligible, suggesting that for $n \geq 16384$, Ozaki scheme II can be performed even more efficiently.

\subsection{Power Efficiency}
\label{subsec:Power Efficiency}
\begin{figure*}[htb]
  \centering
  \includegraphics[width=\hsize]{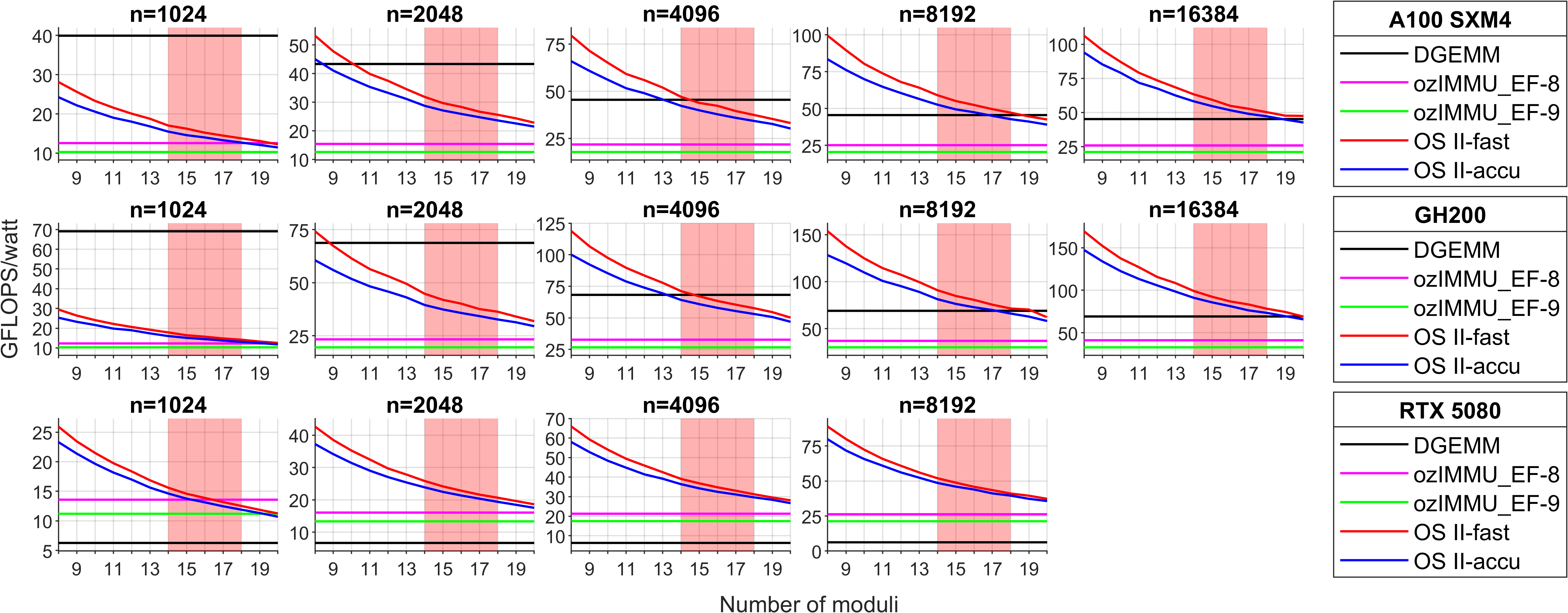}
  \caption{Power efficiency of DGEMM emulation on A100 (top), GH200 (middle), and RTX 5080 (bottom)\label{fig:power-DGEMM}}
\end{figure*}

\begin{figure*}[htb]
  \centering
  \includegraphics[width=\hsize]{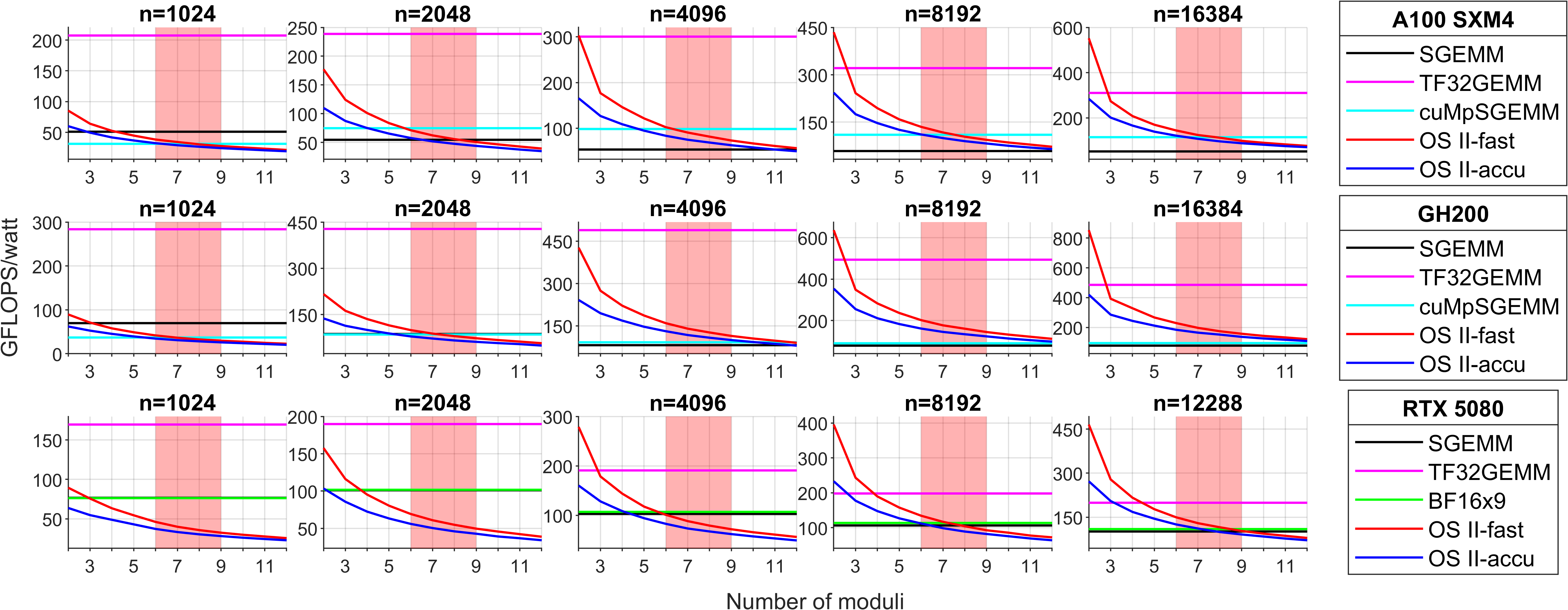}
  \caption{Power efficiency of SGEMM emulation on A100 (top), GH200 (middle), and RTX 5080 (bottom)\label{fig:power-SGEMM}}
\end{figure*}

Figures~\ref{fig:power-DGEMM} and \ref{fig:power-SGEMM} show the power efficiency (GFLOPS/watt) of emulation for $m = n = k$ on RTX 5080 and GH200.
Similar trends were observed on A100 as on GH200.
The power efficiency exhibits trends similar to those of throughput performance, as shown in Figures~\ref{tab:performance-DGEMM} and \ref{tab:performance-SGEMM}.
However, for smaller problem sizes, the results of Ozaki scheme II reached those of existing emulation, DGEMM, and SGEMM.
This can be attributed to the relatively high power efficiency of INT8 GEMM even for moderately sized problems.
For instance, the performance ratio between INT8 GEMM and SGEMM at $n=1024$ was 5.3x, while the power efficiency ratio was as high as 13.3x on RTX 5080.
On GH200, OS II-fast-$N$ and OS II-accu-$N$ achieved 20\%--43\% and 10\%--32\% improvements, respectively, compared to DGEMM for $N \in \{14,15,16,17\}$ and $n = 16384$.
OS II-fast-$N$ with $N \in \{7,8,9\}$ and OS II-accu-$N$ with $N \in \{6,7,8\}$ achieved 103\%--154\% and 96\%--137\% improvements, respectively, compared to SGEMM for $n = 16384$.

\section{Conclusion}
\label{sec:Conclusion}
This paper presented DGEMM and SGEMM emulation based on Ozaki scheme II.
In many cases, applications employing DGEMM do not require the full precision of FP64 (see, e.g.,~\cite{Dawson2024Reducing}).
Ozaki scheme I and II can deliver the required accuracy at a reasonable speed.
Numerical experiments compared native GEMM routines and the proposed emulation in terms of accuracy, throughput performance, and power efficiency.
For sufficiently large problems, the proposed emulation outperformed native DGEMM, native SGEMM, and existing emulation methods in both performance and power efficiency.
In addition, Ozaki scheme II can serve as an intermediate-precision approach between FP32 and TF32.
It can also be extended to matrix multiplication using arbitrary combinations of floating-point formats, including both homogeneous (e.g., double-double) and heterogeneous (e.g., FP16 and FP32) types.
The proposed methods help bridge the gap between low-precision computing units optimized for AI and numerical computations that requires high precision.

\begin{acks}
We appreciate the helpful comments on cuMpSGEMM from Dr. Hiroyuki Ootomo.
We thank Patrick Gutsche, Prajval Kumar, and Dr. William Dawson for their assistance with debugging the code.
This study was supported by Japan Society for the Promotion of Science Grant-in-Aid Numbers 25H01109 (Scientific Research A), 23K28100, 25K03126 (Scientific Research B), and 24K23874 (Research Activity Start-up).
\end{acks}

\bibliographystyle{ACM-Reference-Format}
\input{main.bbl}










\end{document}